\documentclass{PoS}
\usepackage{dcolumn}
\usepackage{graphicx}
\usepackage{tabularx}
\usepackage{slashed}
\usepackage{booktabs}
\usepackage{amsmath}
\def\LL{\left\langle}	
\def\RR{\right\rangle}	
\def\PAR#1#2{ {\frac{\partial #1}{\partial #2}} }
\def\PARTWO#1#2{ {{\partial^2 #1}\over{\partial #2}^2} }

\newcommand{\BE}{\begin{displaymath}}
\newcommand{\EE}{\end{displaymath}}
\newcommand{\BNE}{\begin{equation}}
\newcommand{\ENE}{\end{equation}}
\newcommand{\BEA}{\begin{eqnarray}}
\newcommand{\EEA}{\nonumber\end{eqnarray}}

\newcommand{\Tr}{{\rm Tr~}}

\newcommand{\etal}{{\emph et al.}}

\title{Update on the sea contributions to hadron polarizabilities via reweighting}

\ShortTitle{Sea Polarizability via Reweighting}

\author{\speaker{Walter Freeman}\\
  The George Washington University\\
E-mail: \email{wfreeman@gwu.edu}}

\author{Andrei Alexandru\\
  The George Washington University\\
E-mail: \email{aalexan@gwu.edu}}

\author{Frank X. Lee\\
  The George Washington University\\
E-mail: \email{fxlee@gwu.edu}}

\author{Michael Lujan\\
  The George Washington University\\
E-mail: \email{mlujan@gwu.edu}}

\FullConference{The 2014 Symposium on Lattice Gauge Theory in New York City, USA}

\abstract{
  We have made significant progress on extending lattice QCD calculation of the polarizability of the neutron and other hadrons
  to include the effects of charged dynamical quarks. 
  This is done by perturbatively reweighting the charges of the sea
  quarks to couple them to the background field. 
  The dominant challenge in such a calculation is stochastic
  estimation of the weight factors, and we discuss the difficulties in this estimation. 
  Here we use an extremely aggressive dilution scheme with $N=124,416$ sources
  per configuration to reduce the stochastic noise to a manageable level. We find that $\alpha_E = 2.70(55) \times 10^{-4}\rm{fm}^3$ for the neutron
  on one ensemble. We show that low-mode substitution can be used in tandem with dilution to construct 
  an even better estimator, and introduce the offdiagonal matrix element mapping technique for predicting estimator quality.
}

\begin{document}

\section{Introduction}

The GWU lattice QCD group is engaged in an extended program to compute hadron electric polarizabilities with high precision
and at the physical point. We use the background field method, in which the energy of the hadronic ground state is measured at zero and nonzero
electric field $\bf \mathcal E$; the polarizability is related to the energy shift by $\Delta E = \frac{1}{2}\alpha \bf \mathcal E^2$. 
This technique is well-understood,
and the challenge in its application lies in bringing it in contact with the physical point.
An extremely difficult
part of our progress toward the physical point lies in charging the sea quarks. We accomplish this {\it via} perturbative reweighting,
and this work addresses the details of this technique and the results obtained from it.

This proceedings is organized as follows. First, we briefly outline the background field method and perturbative reweighting, techniques explained in greater 
detail elsewhere. We then focus on the most difficult aspect of the calculation: stochastic estimation of the weight factors. We first discuss why the stochastic
estimators involved here are so badly behaved, then outline the use of dilution as an improvement technique. 

We have completed a calculation on the $24^3 \times 48$ ensemble using a $N=124,416$ dilution scheme for the reweighting factor stochastic estimators,
and present the results in Sec.~\ref{sec_2448}. We would like to further reduce the statistical errors, and doing so while applying the same technique
to our larger or more chiral ensembles would require significant extra computing power. Thus, we explore the use of low-mode substitution (LMS) in tandem
with dilution to achieve a lower statistical error at a reduced cost; we outline this technique and describe an ongoing run on the $48 \times 24^2 \times 48$ ensemble
in Sec.~\ref{sec_482448}.
%

The broader GWU polarizability project uses a series of dynamical ensembles using $n_f=2$ flavors of nHYP-smeared Wilson-clover 
fermions~\cite{Hasenfratz07} at two different pion masses and a variety of volumes with $a=0.1215$ fm; each has 300 configurations.
While we ultimately want to measure the hadron polarizabilities on all of these ensembles, in this work
we present results from only two. The completed run was done on a $24^3 \times 48$ ensemble; the run still in progress using LMS
uses a $48 \times 24^2 \times 48$ ensemble, with the electric field pointing along the elongated direction.
For more details on these ensembles, see \cite{MikesPRD}.

\section{Methods}
\label{sec_methods}
Here we give only a brief overview of the background field method and our simulations. We refer the reader to \cite{MikesPRD} for a detailed description of the methods
and results from the valence-only calculation. We apply the electric field by adding a U(1) phase to the gauge links, and 
avoid consistency issues by choosing Dirichlet boundary conditions in the $x$ and $t$ directions. 
We then measure correlators with and without the electric field to determine the energy shift.
Because the correlators are strongly correlated with each other we can determine $\Delta E$ to much better absolute precision than $E$ itself.
The parameters are extracted by a fit to
%
  $G(t,\eta) = (A + \eta^2 B) e^{-(m + \eta^2 C)t}$,
%
where the parameter $C$ is related to the polarizability.

However, computing the valence correlators using a set of gauge links with U(1) phase factors applied only couples the electric field to the valence quarks.
A $\chi$PT calculation suggests that 20\% of the neutron polarizability comes from the contribution of the charged sea\cite{Detmold06}, 
so it is essential to include
this element in any physically-meaningful lattice computation of hadron polarizabilities. 
We could in principle generate a second ensemble with a background electric field and compute the $\eta \neq 0$ correlators on it. However, doing so
eliminates the strong correlations we rely on to achieve a small error in $\Delta E$. What we need are two ensembles with different sea quark dynamics but 
which are correlated; this can be achieved {\em via} reweighting.

Conventional reweighting requires the stochastic estimation of a determinant ratio to obtain the weight factors. 
While there are several techniques that can produce
reasonable results for the $w$'s for mass reweighting\cite{Hasenfratz08}, the stochastic estimation of the weight factors is the 
chief difficulty in this calculation. The standard techniques used for mass reweighting estimators fail here. Thus, we
turn to a technique we call {\em perturbative reweighting}. We only outline this technique here; for a more detailed explanation see \cite{Freeman:2013eta}. The idea is to estimate $\PAR{w}{\eta}$, the derivatives of the weight factor with respect to the 
electric field strength, rather than $w$ itself. 
Since we are interested only in a perturbatively-small shift in the action, we may then expand the weight factor in powers of $\eta$, keeping terms up to
second order as we are looking for a quadratic effect. 
%
%
The derivatives $w'=\PAR{w}{\eta}$ and $w''=\PARTWO{w}{\eta}$ can be expressed as the traces of operators using Grassman integration; we obtain

\begin{equation}
  w' = \PAR{}{\eta} \frac{\det M_\eta}{\det M}=
\Tr \left( M' M^{-1} \right) \,,
\label{eq:9}
\end{equation}
and
\begin{align}
  w''= \PARTWO{}{\eta} \frac{\det M_\eta}{\det M} = \Tr \left( M'' M^{-1} \right) - \left(\Tr M' M^{-1} \right)^2 - \Tr \left(M' M^{-1} \right)^2 \,,
\label{eq:10}
\end{align}
where $M'$ and $M''$ are the derivatives with respect to $\eta$ at $\eta=0$
of the one-flavor fermionic matrix $M$. 
In order to determine $w'$ and $w''$, we need estimates of $\Tr M' M^{-1}$, $\Tr M' M^{-1}  M' M^{-1}$, $\Tr M'' M^{-1}$, and $\left( \Tr  M' M^{-1} \right)^2$; two 
independent estimates of the first can be combined to estimate the last. We refer to the combination of the middle two terms which must be estimated as $\tilde w'' = \Tr \left( M'' M^{-1} - M' M^{-1}  M' M^{-1} \right)$,
as these nearly cancel.

To obtain the two-flavor weight factor $w$ at some particular value of $\bf \mathcal E$
corresponding to $\eta_d$ for the down quark and $-2\eta_d$ for the up quark,
we multiply the single-flavor weight factors $w_1$ for the two flavors, keeping terms only up to order $\eta^2$.
\noindent The valence correlators with nonzero field $\LL G_\eta(t)\RR_0$ consider only the charges of the valence quarks.
With the two-flavor weight factors in hand, we may reweight them in the usual way to generate fully-charged correlators $\LL G_\eta(t)\RR_\eta = \frac{\sum w(\eta_d) G_\eta(t)}{\sum w(\eta_d)}$, where the sums
run over configurations. The resulting fully-charged correlators are then used to compute the polarizabilities as in the valence-only background field method.

\section{Estimating the weight factor coefficients: body-centered hypercubic dilution}

We now turn to the most difficult aspect of this calculation: computing stochastic estimates of $w'=\Tr M'M^{-1}$ and $w''=\Tr M''M^{-1} - \left( M'M^{-1} \right)^2$.
We adopt the conventional estimator using Z(4) noise vectors $\xi$ for the trace. This estimator's variance is given by

\begin{equation}
  \rm{var} (\xi^\dagger \mathcal O \xi) = \sum_{i \neq j} \left| \mathcal O_{ij} \right|^2
\end{equation}

\noindent where the sums run over both spatial and spin/color indices. Thus, the variance depends greatly on the offdiagonal structure of $\mathcal O$. Certain operators (notably, $M^{-1}$) are diagonally-dominant and their estimators converge quickly. 
We are not so fortunate here; the operator $M' M^{-1}$, for instance, corresponds to $M^{-1}$ with a point-split current; this shifts the large diagonal elements of $M^{-1}$ just off the diagonal, 
suppressing the trace (our desired signal) and enhancing the stochastic noise. 

One common technique for reducing the stochastic noise is dilution, in which the dimension of $\mathcal O$ is split into $N$ subspaces and the trace over each estimated separately. 
This is done by constructing $N$ noise vectors $\xi$, each of which has support on only one subspace. If we label the subspace that a lattice site $i$ belongs to as $p(i)$, then the variance becomes

\begin{equation}
  \rm{var} (\xi^\dagger \mathcal O \xi) = \sum_{i \neq j} \delta_{p(i) p(j)} \left| \mathcal O_{ij} \right|^2,
\end{equation}
 
\noindent at the cost of requiring $N$ evaluations of the operator in question. This same effort could be used to reduce the noise by a factor of $N$ by simple repetition; whether dilution pays off depends on whether the
elements which still contribute to the noise ({\em i.e.} those with $p(i) = p(j)$) are substantially smaller than those which are removed.
It is thus informative to map the $\mathcal O_{ij}$'s in order to design a dilution scheme or to evaluate other estimator improvement techniques.

\begin{figure}[ht]
  \begin{center}
    \includegraphics[width=0.4\textwidth]{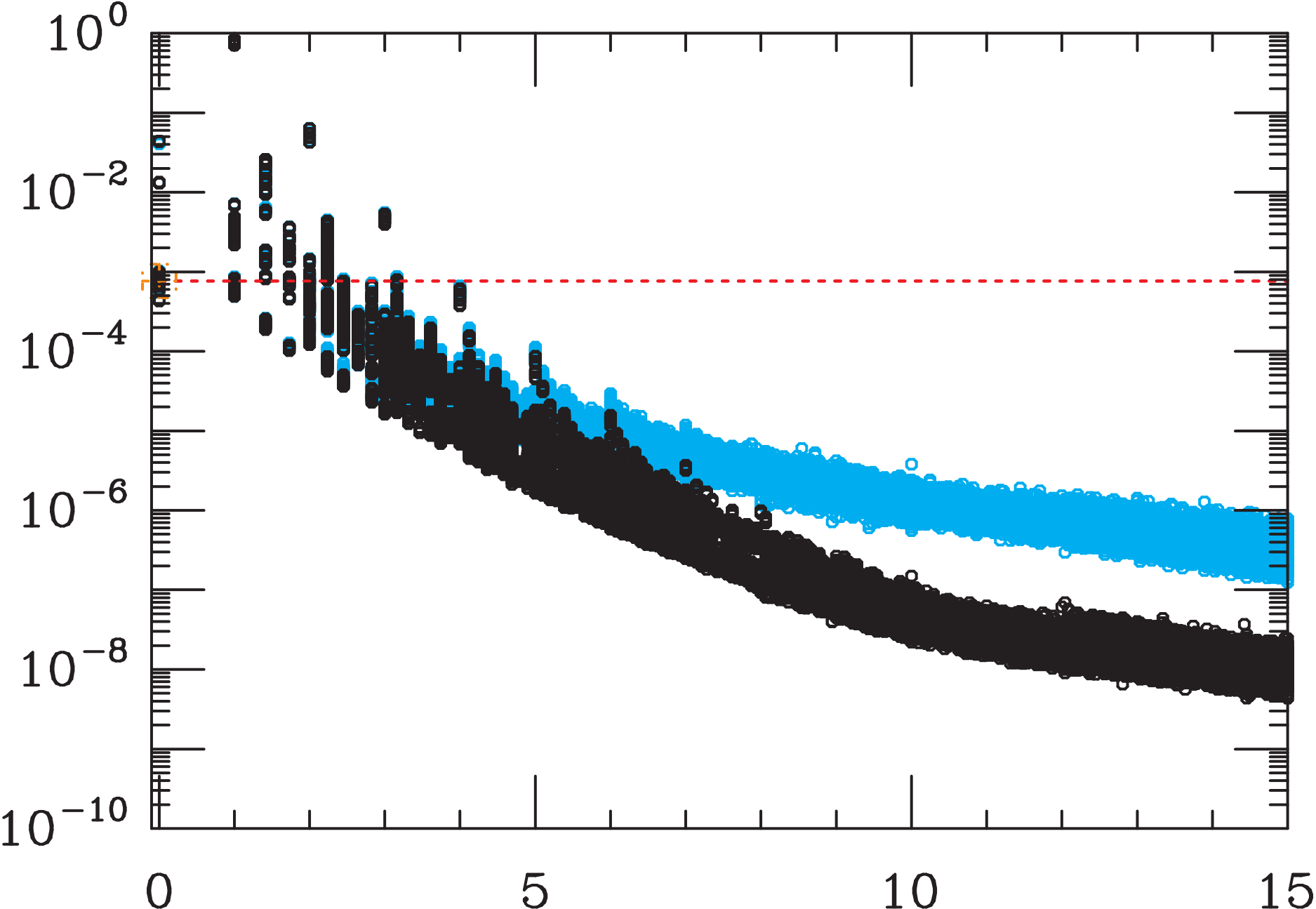}
    \hspace{0.1\textwidth}
    \includegraphics[width=0.4\textwidth]{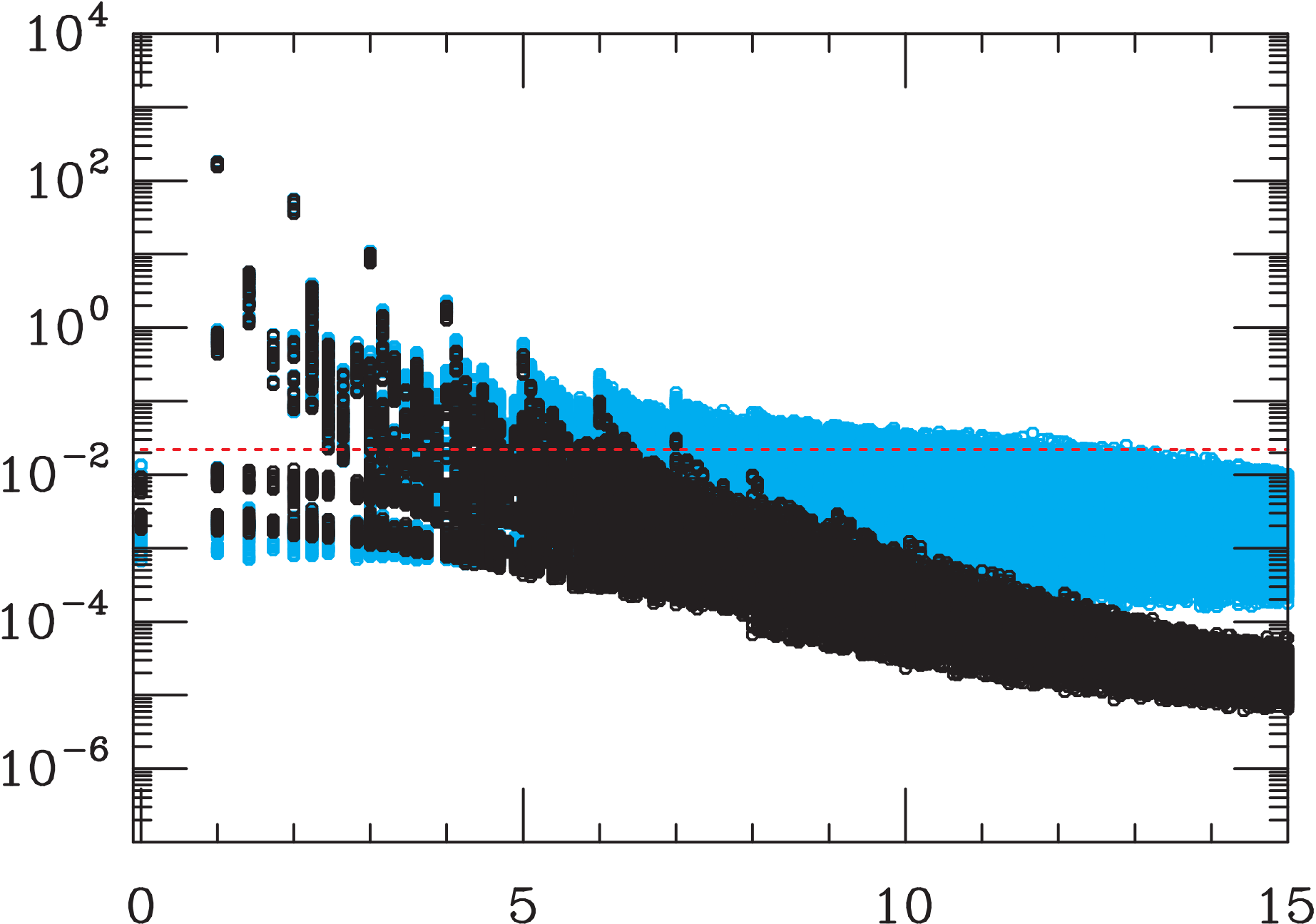}
  \end{center}
    \caption{A mapping of the mean offdiagonal elements of $w'$ (left) and $\tilde w''$ (right) as a function of the Euclidean separation between source and sink.
      The light blue points show the unimproved values of the operators. The black points show the effect of low-mode subtraction with 500 eigenvectors of 
    $\gamma_5 M$. (The unimproved points lie under the LMS-improved points for short distances.) The red horizontal line shows the mean squared value of the diagonal
  elements for comparison.}
  \label{fig-lms-offdiag}
\end{figure}

We have done this by computing all $\mathcal O_{ij}$ for a set of 72 sources $j$ spread across the lattice for one configuration, and binning together those elements $\mathcal O_{ij}$ that are related by reflection, rotation, and translation symmetries in the 
gauge average. We find that the size of the $\mathcal O_{ij}$'s
depends primarily on the Euclidean distance separating $i$ and $j$, with temporal separation somewhat more important than spatial separation for both $w'$ and $\tilde w''$. This is consistent with the interpretation of $M'$ as a point-split current
in the temporal direction. The dependence of $w'_{ij}$ and $\tilde w''_{ij}$ on this Euclidean separation is shown in Fig. \ref{fig-lms-offdiag}, along with the contribution to the overall variance as a function of separation. As expected, the near-diagonal elements are quite large,
larger than the diagonal elements which comprise the signal. 

A good
dilution scheme is thus one that maximizes the distance between nearest neighbors in the same subspace.
One such scheme uses subspaces consisting of a body-centered hypercubic (BCHC) grid. Each BCHC
subspace with unit cell $2 \Delta$ consists of the standard grid augmented by another copy offset by a shift $(\Delta, \Delta, \Delta, \Delta)$.
As noted previously the offdiagonal elements are somewhat larger if the separation between $i$ and $j$ is more timelike. It is possible to leverage this by creating an anisotropic BCHC dilution scheme with $\Delta_t > \Delta_{xyz}$; we will use such a scheme
for the $48 \times 24^2 \times 48$ ensemble.

We may also use the offdiagonal element mapping data to estimate the stochastic variance of any proposed dilution scheme.
More aggressive dilution schemes outperform simple repetition for $w'$, but feasible ones do not yet show large gains 
over repetition of a na\"{i}ve estimator for $\tilde w''$.


\section{The first ensemble}
\label{sec_2448}
\begin{figure}
  \begin{center}
\includegraphics[width=0.365\textwidth]{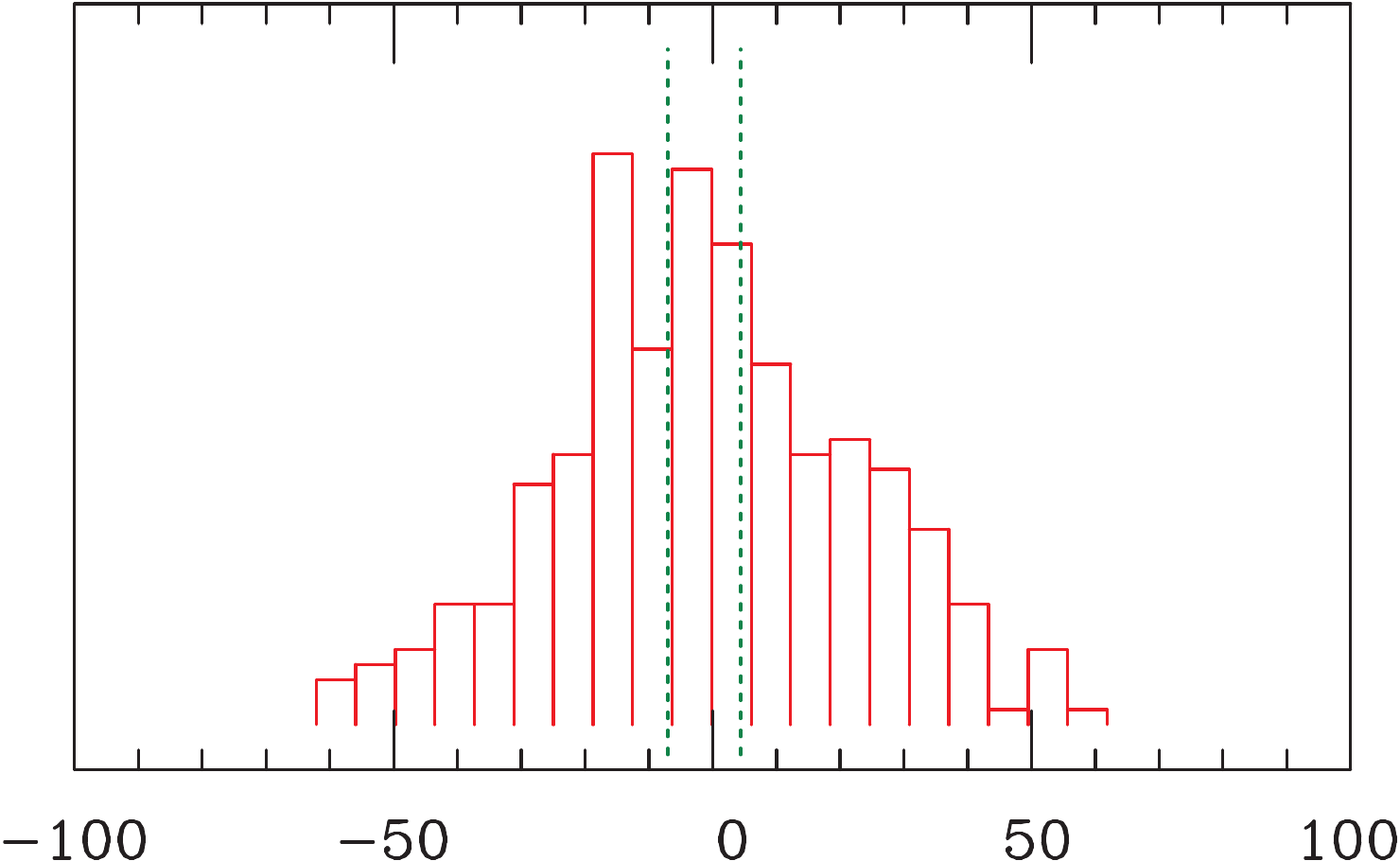}
\hspace{0.1\textwidth}
\includegraphics[width=0.4\textwidth]{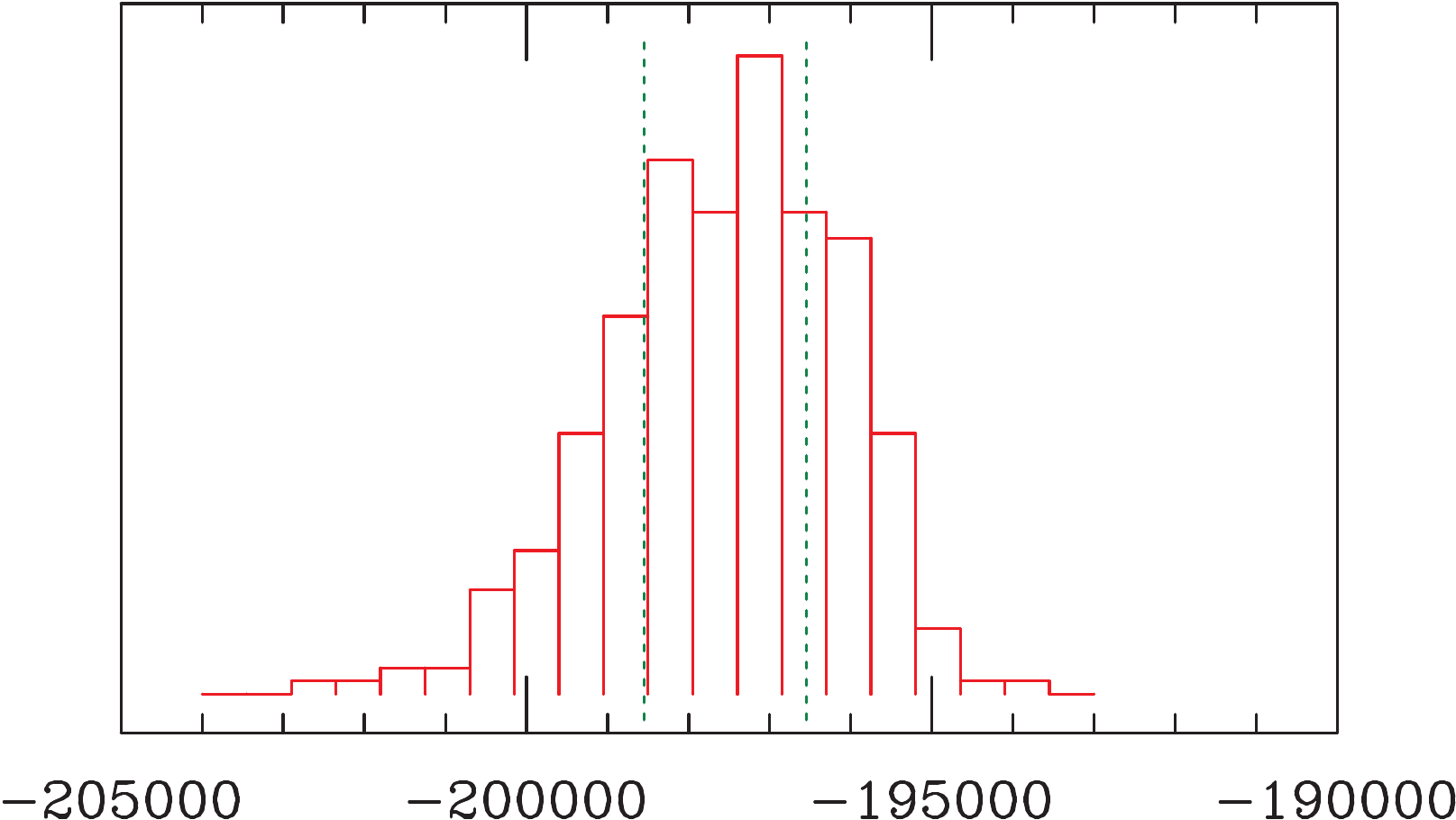}
\end{center}
\caption{Estimates of $\Tr w'$ (left) and $\Tr \tilde w''$ (right) on the $24^3\times48$ ensemble, computed using the $N=124,416$ $\Delta=6$ BCHC dilution scheme. The green dashed lines show the mean value and the error on the stochastic estimator.}
\label{fig-weight-derivs}
\end{figure}

This full analysis of estimator variance was not completed when we embarked upon this run. In particular, we used the estimation of $\Tr w'$ as our metric in estimator design; as the preceding analysis shows, estimation of $\Tr \tilde w''$ is more difficult 
still and is the larger driver of the final error bar. We chose to use a $\Delta=6$ BCHC + spin/color dilution scheme; this scheme has $124,416$ subspaces. 
The first two panels of Fig.~\ref{fig-weight-derivs} show the distribution of $w'$ and $\tilde w''$ along with the stochastic error in each estimate; we note that the stochastic error is quite low compared to the gauge fluctuation (the signal) for $w'$, while
the two are comparable for $\tilde w''$. 
The additional term appearing at second order, $(\Tr M' M^{-1})^2$, presents another hurdle, as it requires two independent estimates of $\Tr M' M^{-1}$. 
As 
this term contributes rather little to the total value of $w''$ we use a second lower-quality estimate obtained from a trial of another technique.
With $w'$ and $\tilde w''$ computed on each configuration, we determine the weight factor $w$ to second order in $\eta_d$ for our chosen value $\eta_d=10^{-4}$. 
We then compute the polarizabilities of the neutron, neutral kaon, and ``neutral pion''\footnote{The neutral pion studied here does not include the disconnected contributions of the physical $\pi^0$.}
with a correlated fit between the zero-field correlators and the (reweighted) nonzero-field correlators
as in the valence-only case, described in Section~\ref{sec_methods} and more fully in Ref.~\cite{MikesPRD}. The results are shown in Table~\ref{table-results} with the sea effects ``turned on'' order by order: valence only,
first-order sea effects (including the $w'$ term only), second-order with $\tilde w''$ only, and second order with both contributions. 
Despite the large increase in the uncertainty
from the inclusion of the sea effects, the overall error bar is small compared to other polarizability calculations. The central values do not shift much, with the curious exception of the $K^0$; this is consistent with
the findings in~\cite{MikesPRD}, where the $K^0$ polarizability showed a large sensitivity to the sea quark mass. Of all the ensembles in our polarizability study, this one is expected to show the smallest effect from
the charging of the sea, as it has the larger of the pion masses and the smallest volume.

\begin{table*}[t]
  \scriptsize
  \begin{tabular*}{0.95\textwidth}{@{\extracolsep{\stretch{1}}}l*{13}{r}@{}}
    \toprule
            &\phantom{ab}& \multicolumn{2}{c}{Valence only} &\phantom{a}&
            \multicolumn{2}{c}{$1^\text{st}$ order} &\phantom{a}& \multicolumn{2}{c}{$\tilde{w}_q''$ only}
            &\phantom{a}& \multicolumn{2}{c}{$2^\text{nd}$ order} \\
    \cmidrule{3-4}\cmidrule{6-7}\cmidrule{9-10}\cmidrule{12-13}
  &&\multicolumn{1}{c}{$a\Delta E$}          & \multicolumn{1}{c}{$Q$}
            && \multicolumn{1}{c}{$a\Delta E$}             & \multicolumn{1}{c}{$Q$}
            && \multicolumn{1}{c}{$a\Delta E$}              & \multicolumn{1}{c}{$Q$}
            && \multicolumn{1}{c}{$a\Delta E$}              & \multicolumn{1}{c}{$Q$}       \\
    \midrule
    Pion    &&  -5.4(3.4)   & 0.17  &&  -6.0(3.4)     & 0.18     &&  5.4(5.6)       & 0.15     &&  5.6(5.7)       & 0.15    \\
    Kaon    &&  4.2(0.8)    & 0.12  &&  3.7(1.0)      & 0.07     &&  10.5(3.4)      & 0.03     &&  11.1(3.4)      & 0.02    \\
    Neutron &&  62.8(5.7)   & 0.65  &&  63.9(6.5)     & 0.57     &&  72.5(16.4)     & 0.53     &&  67.0(16.3)     & 0.43    \\
    \bottomrule
  \end{tabular*}
  \caption{Results for the energy shift for the $\pi^0$, neutron, and $K^0$ with differing orders
    of reweighting: none (the valence-only calculation), first-order in $\eta_d$, second-order
    including only the contribution from $\tilde{w}''$, and the full calculation to
    second order. For the energy shifts, the values are in units of $10^{-8}$. $Q$ is the confidence
   level for the fits corrected to account for the sample size~\cite{Toussaint:2008ke}.}
  \label{table-results}
\end{table*}

\section{Low-mode subtraction as a further improvement technique}
\label{sec_482448}
We now turn to the $48 \times 24^2 \times 48$ ensemble and the low-mode subtraction technique we are using to study it.
The results from the previous section illustrate that the dilution technique is successful at improving the estimator for $w'$, but that further improvement is still
useful for $\tilde w''$, especially for the other costlier ensembles. Dilution is successful for the first-order term because of the exponential falloff with Euclidean separation shown in 
\ref{fig-lms-offdiag}; it is less successful at second order because the falloff is slower. The rate of this falloff is governed essentially by the 
value of $m_\pi$. We note that the long-distance behavior of the pion correlator is well-saturated by the low-lying eigenmodes of $\gamma_5 M$ with Wilson
fermions~\cite{Neff}. Thus, we may remove the low modes of $\gamma_5 M$ from $M^{-1}$ to accelerate this falloff and gain more benefit from dilution. 
We compute the trace over the high sector stochastically as before; the trace over the low sector can be written as a sum over eigenvalues and evaluated
exactly.

\begin{figure}[t]
  \begin{center}
 \includegraphics[width=0.385\textwidth]{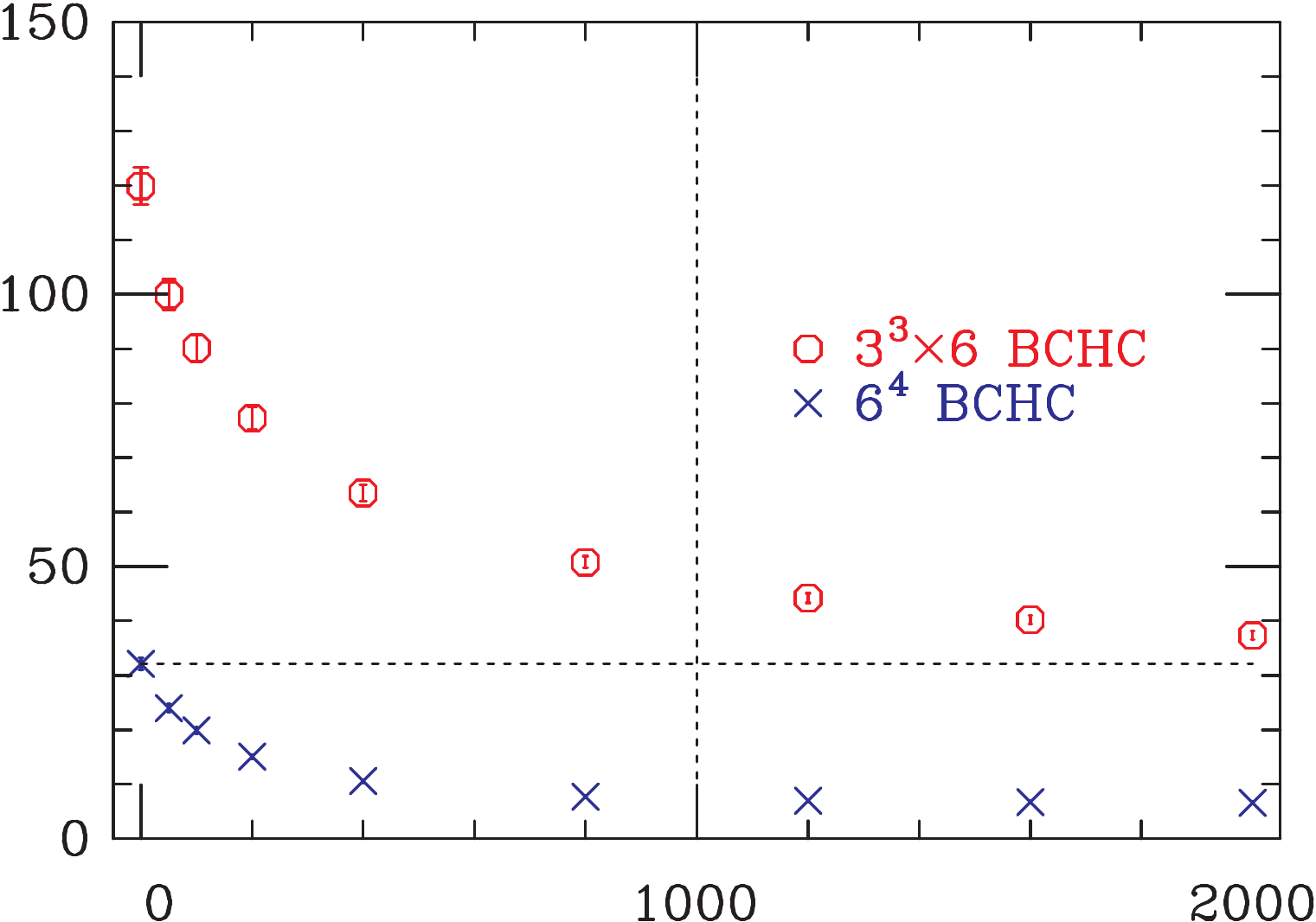}
 \hspace{0.1\textwidth}
  \includegraphics[width=0.4\textwidth]{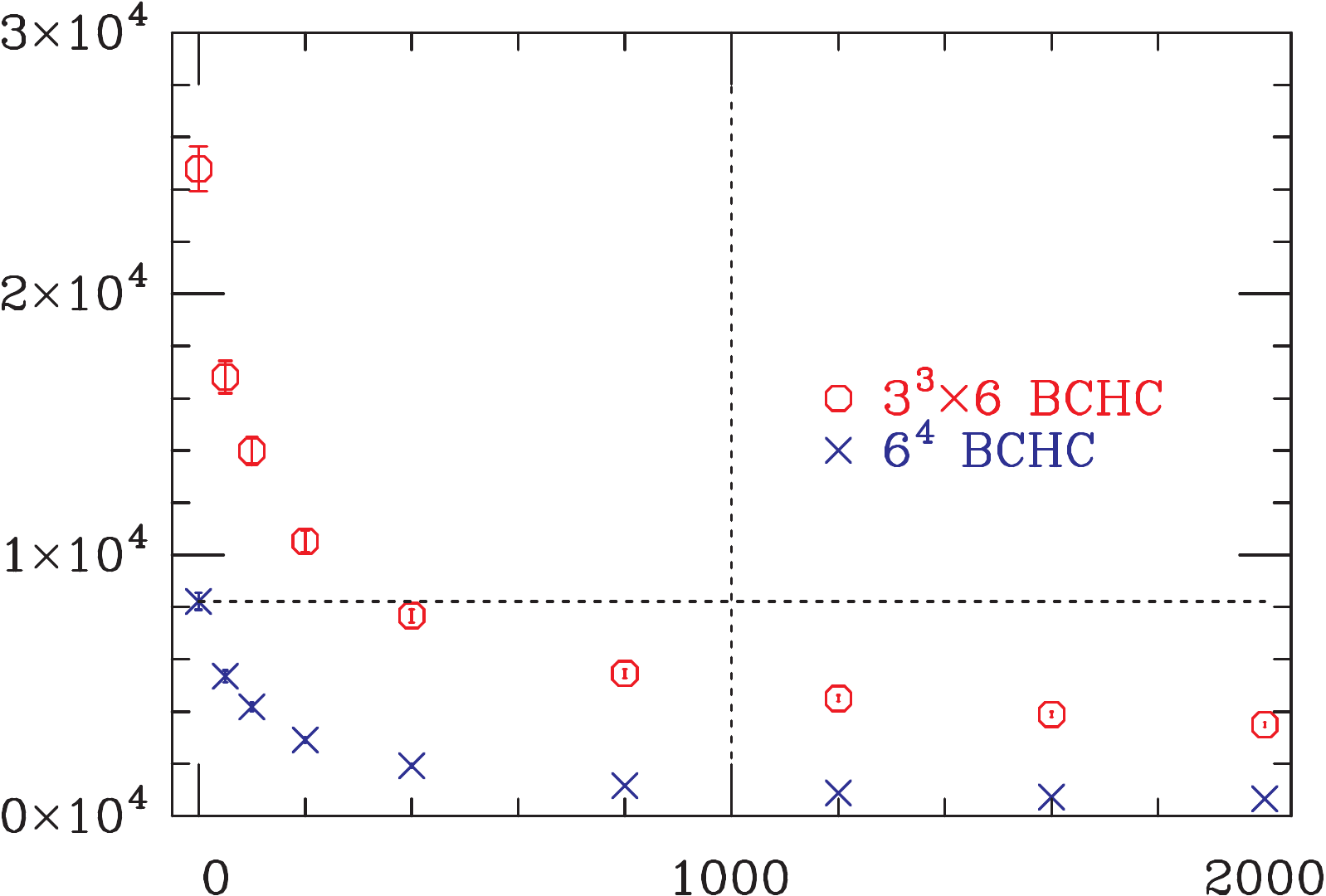}
\end{center}
  \caption{Stochastic error in the estimator of $w'$ (left) and $\tilde w''$ (right) vs. the number of eigenvectors used for LMS, using two different BCHC
    dilution schemes on the $48 \times 24^2 \times 48$ ensemble. 
    The vertical line shows the 1000-vector eigensystem size we decided gave the best cost-performance tradeoff, while the horizontal line indicates
    the performance of the $6^4 BCHC$ dilution scheme without LMS for comparison. Note that the 1000-vector LMS applied to the $3^3 \times 6$ dilution scheme, requiring only
  one-eighth the number of subspaces, outperforms the $6^4$ scheme with no LMS for $\tilde w''$.}
  \label{fig-perf}
\end{figure}

To determine whether low mode subtraction will be effective at reducing the stochastic noise, we return to the offdiagonal matrix element mapping technique. Fig. \ref{fig-lms-offdiag} shows the offdiagonal matrix elements for 500 eigenvectors, with the unimproved
operator shown for comparison; the improvement is clear. We continue to see further reductions in the estimated stochastic variance up through 2000 eigenvectors, the largest eigensystem we tried; in planning a production run, we must balance the 
improvement with the extra cost and logistical burden of handling larger eigensystems. 
Fig.~\ref{fig-perf} shows the tradeoff between eigenspace size and stochastic error reduction for two different dilution schemes, the $6^4$ scheme used for the 
$24^3 \times 48$ ensemble without LMS, and the $3^3 \times 6$ scheme we plan to use for the $48 \times 24^2 \times 48$ ensemble. We note that the overall error 
for the latter ensemble is substantially larger than might be expected. We believe this is due to a gauge-dependent contribution to the stochastic noise. We will address
this point further in a future publication.

Based on the data in Fig.~\ref{fig-perf} we have 
elected for an anisotropic $3^3 \times 6$ BCHC dilution scheme ($N_{\rm{dil}} = 15552$) along with LMS using 1000 eigenvectors, outperforming
the $6^4$ scheme at a fraction of the cost.
The dominant contribution to the computational cost is still the stochastic estimates; the eigensolver and exact traces are comparatively cheap.

\section{Conclusion}

We have developed techniques for using strong dilution in combination with LMS to estimate the perturbative reweighting coefficients associated with charging
the quark sea. Even without LMS, our techniques have allowed us to compute the most precise value of the neutron polarizability including these effects, 
$\alpha_E = 2.70(55)\times 10^{-4}\textrm{fm}^3$, on one of our ensembles. In order to carry out an infinite-volume and chiral extrapolation, we need similar results from the
other ensembles with different sizes and $m_\pi$; the first of these runs using LMS with 1000 eigenvectors is underway.

\section{Acknowledgements}

This calculation was done on the GWU Colonial One and IMPACT GPU clusters and the Fermilab GPU cluster. This work is
supported in part by the NSF CAREER grant PHY-1151648 and the U.S. Department of Energy grant DE-
FG02-95ER-40907.

    \end{document}